\documentclass[A4]{aa}
\usepackage{graphicx}

\begin{document}

\title{GRB021125: the first GRB imaged by INTEGRAL\thanks{Based on observations with
INTEGRAL, an ESA project with instruments and science data centre funded by ESA member
states (especially the PI countries: Denmark, France, Germany, Italy, Spain, Switzerland),
Czech Republic and Poland, and with the participation of Russia and the USA.}}

\author{G. Malaguti\inst{1}, A. Bazzano\inst{2}, V. Beckmann\inst{3,4}, A.J. Bird\inst{5},
M. Del Santo\inst{2}, G. Di Cocco\inst{1}, L. Foschini\inst{1}, P. Goldoni\inst{6},
D. G\"otz\inst{7}, S. Mereghetti\inst{7}, A. Paizis\inst{3}, A. Segreto\inst{8},
G. Skinner\inst{9}, P. Ubertini\inst{2} \and A. von Kienlin\inst{10}}

\offprints{G. Malaguti \email{malaguti@bo.iasf.cnr.it}}

\institute{Istituto di Astrofisica Spaziale e Fisica Cosmica (IASF) del
CNR, Sezione di Bologna, Via Gobetti 101, 40129 Bologna (Italy)
\and
Istituto di Astrofisica Spaziale e Fisica Cosmica (IASF) del
CNR, Sezione di Roma, Via Fosso del Cavaliere 100, 00133 Roma (Italy)
\and
INTEGRAL Science Data Centre, Chemin d'Ecogia 16, CH--1290 Versoix (Switzerland)
\and
Institut f\"ur Astronomie and Astrophysik, Universit\"at T\"ubingen, Sand
1, 72076 T\"ubingen (Germany)
\and
School of Physics and Astronomy, University of Southampton, Highfield, Southampton, SO17 1BJ, UK.
\and
CEA Saclay, DSM/DAPNIA/SAp, F--91191, Gif sur Yvette Cedex (France)
\and
Istituto di Astrofisica Spaziale e Fisica Cosmica (IASF) del
CNR, Sezione di Milano, Via Bassini 15, 20133 Milano (Italy)
\and
Istituto di Astrofisica Spaziale e Fisica Cosmica (IASF) del
CNR, Sezione di Palermo, Via U. La Malfa 153, 90146 Palermo (Italy)
\and
Centre d'Etude Spatiale des Rayonnements, 9 Avenue du Colonel Roche, F--31028,
Toulouse Cedex 4 (France)
\and
Max-Planck-Institut f\"ur Extraterrestrische Physik, Giessenbachstrasse,
85748 Garching (Germany)
}

\date{Received 14 July 2003; accepted 30 July 2003}

\abstract{In the late afternoon of November $25^{\mathrm{th}}$,
2002 a gamma--ray burst (GRB) was detected in the partially coded field
of view (about $7.3^{\circ}$ from the centre)
of the imager IBIS on board the INTEGRAL satellite.
The instruments on-board INTEGRAL allowed, for the first time, the observation of the prompt gamma-ray emission over a broad energy band from 15 to 500 keV.
GRB021125 lasted $\sim24$~s with a mean flux of $\sim5.0$~photons~cm$^{-2}$~s$^{-1}$ in the $20-500$~keV energy band, and a fluence of
$\sim4.8\times 10^{-5}$~erg~cm$^{-2}$ in the same energy band.
Here we report the analysis of the data from the imager IBIS
and the spectrometer SPI.
}

\authorrunning{G. Malaguti et al.}

\maketitle

\section{Introduction}

\begin{figure}
\centering
\includegraphics[angle=270,scale=0.35]{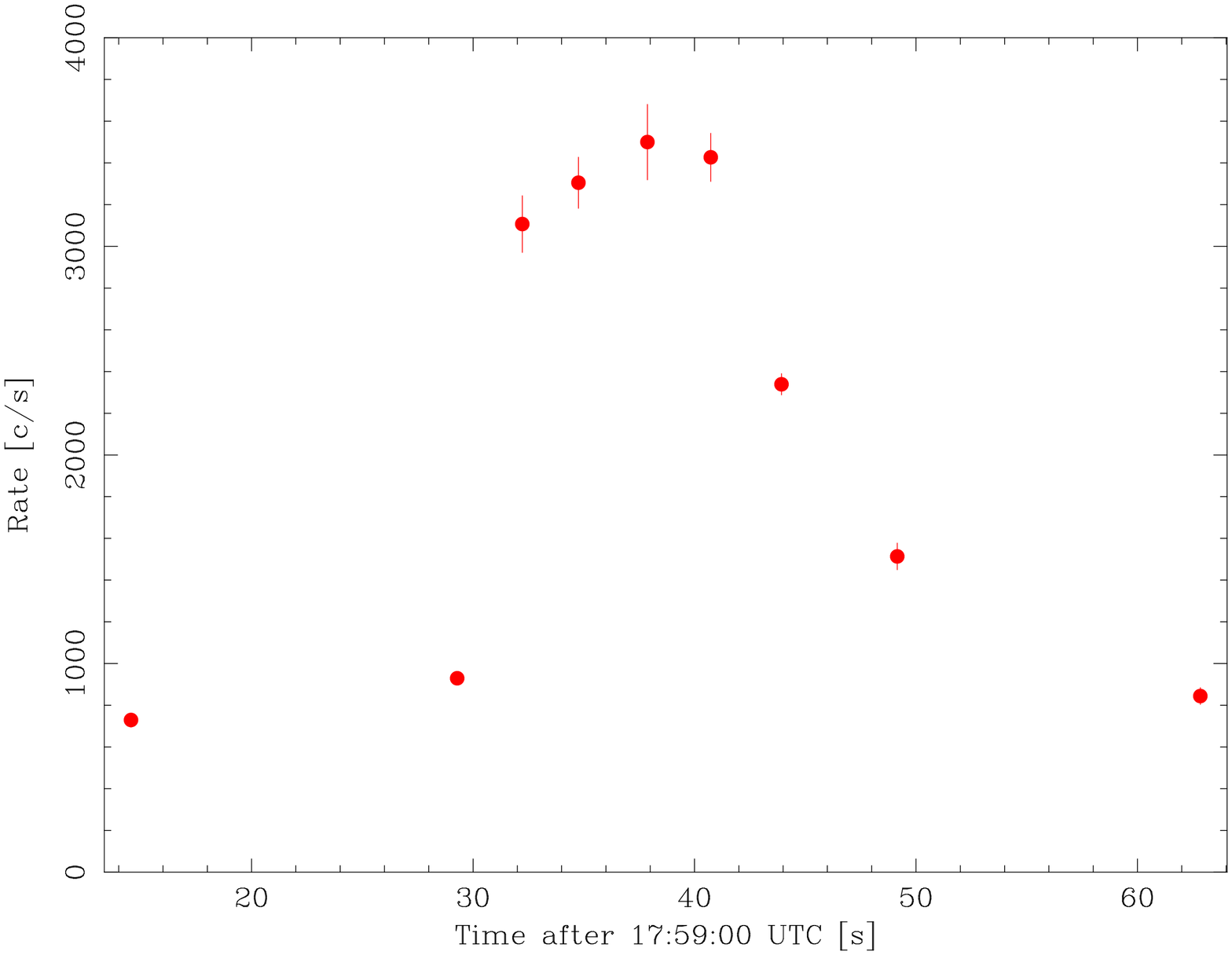}\\
\includegraphics[scale=0.45]{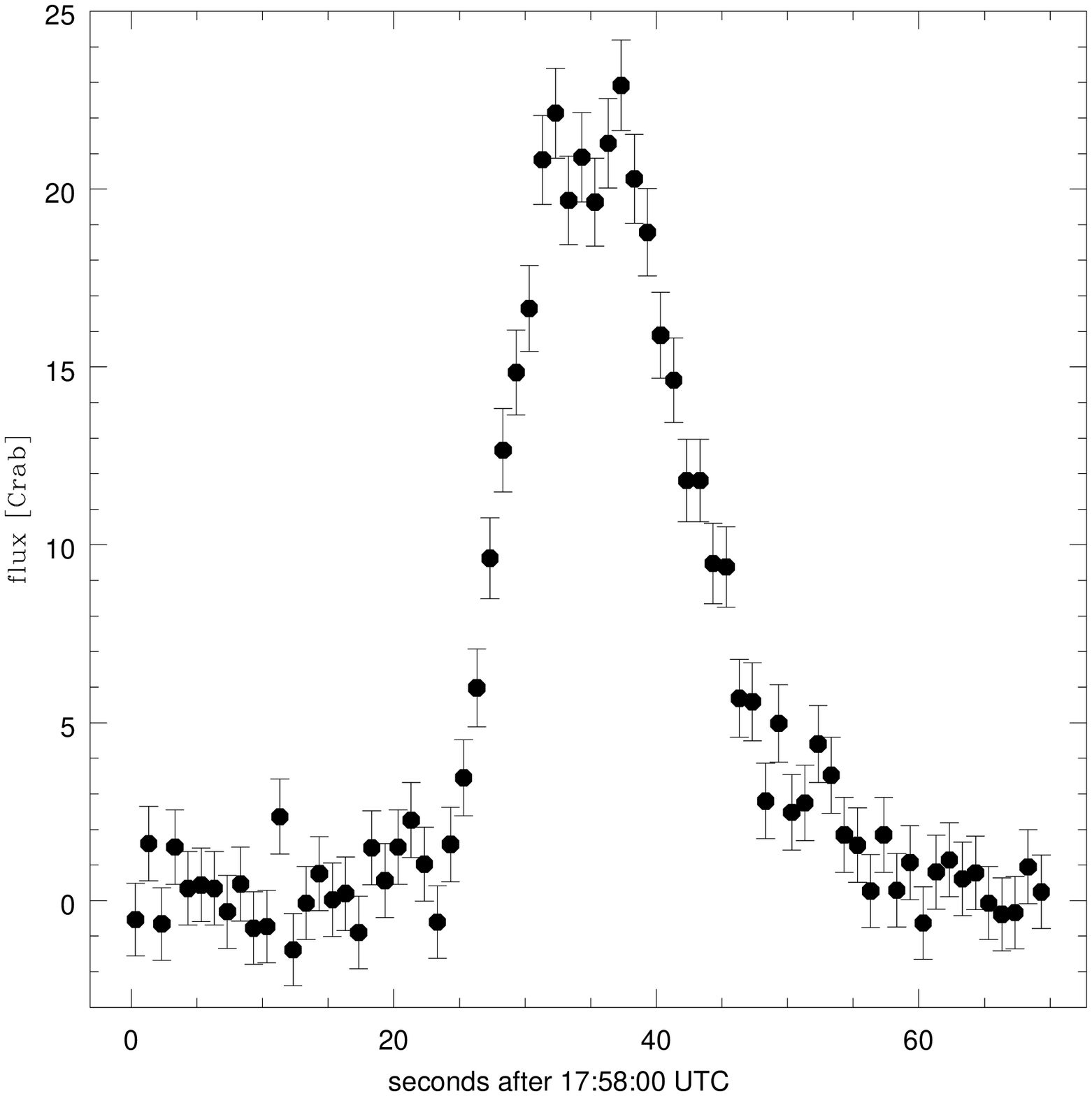}
\caption{GRB021125 lightcurves: ($top$) IBIS/ISGRI lightcurve in the whole energy range,
heavily affected by telemetry gaps; time starts from 17:58:00 UT. ($bottom$)
SPI lightcurve (in Crab units) obtained from the detector count rates in the energy
range $0.02-8$~MeV, which are part of the scientific housekeeping data; time starts from $17:58:00$~UT.}
\label{lcurve1}
\end{figure}

INTEGRAL is the ESA satellite dedicated to the astrophysics
in the X-- and $\gamma$--ray domain, launched on October $17^{\mathrm{th}}$, 2002 (Winkler et al. 2003).
It is composed of two main high--energy telescopes (IBIS, SPI) coupled with two
monitors, one in the X--ray energy band (JEM--X, Lund et al. 2003) and the other
working at optical wavelengths (OMC, Mas--Hesse et al. 2003). IBIS (Ubertini et al. 2003) has
moderate energy resolution, and is optimized for fine imaging, with
$12'$ angular resolution, and $\leq1'$ point source location accuracy for
$\geq 30\sigma$ detections (Gros et al. 2003) in a $9^{\circ}\times 9^{\circ}$
fully coded field of view.
IBIS is composed of two layers: ISGRI (Lebrun et al. 2003) working in the energy band
$15-1000$~keV, and PICsIT (Di Cocco et al. 2003) operating from $175$~keV to $10$~MeV.
The spectrometer SPI provides spectra with high energy resolution ($3$~keV at $1.7$~MeV)
in the energy band from $20$~keV to $8$~MeV (Vedrenne et al. 2003).

Just a few weeks after the launch, the INTEGRAL satellite started the
in--orbit calibration observing Cyg X--1. On November $25^{\mathrm{th}}$, 2002, the satellite was
set up for a special observation with the PICsIT layer in the non--standard photon--by--photon
mode, a reduced number of channels, and most of the satellite telemetry.
This special configuration was required since PICsIT operates in an energy band where
the background rate is very high ($\approx 3500$~counts/s on the whole detector)
and the available telemetry is not sufficient to download all the data (for more details
on the PICsIT modes of operation see Di Cocco et al. 2003) in photon-by-photon.
So, to perform the calibration of PICsIT photon--by--photon mode
it was necessary to limit the operative range at $<500$~keV and to strongly reduce the
telemetry allocation to the other instruments (SPI, JEM--X, OMC were sending only
housekeeping data to ground) and to the ISGRI layer of IBIS.

During this test, at 17:58:30~UTC a gamma--ray burst occurred in
the partially coded field of view of IBIS (about $7.3^{\circ}$ off--axis),
and lasted about $24$~s (Bazzano \& Paizis 2002).
The burst was soon confirmed by the InterPlanetary Network (IPN) composed
of the satellites Ulysses, Mars Odyssey--HEND, and RHESSI (Hurley et al. 2002).
Here we report on the analysis of the data from the INTEGRAL instruments.

\section{Lightcurves}
\begin{figure*}
\centering
\includegraphics[scale=0.5]{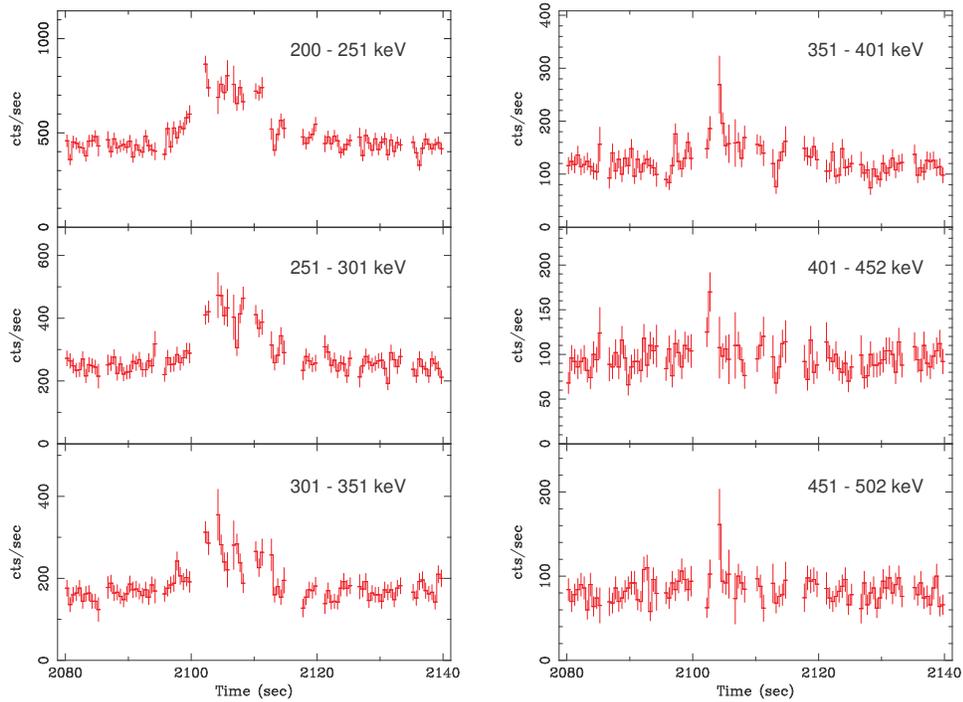}
\caption{GRB021125: IBIS/PICsIT lightcurve for single events in
different energy bands. Time starts from 17:23:50 UT.}
\label{lcurve}
\end{figure*}

Lightcurves were available from IBIS and SPI instruments. For the former, the ISGRI
detector data were strongly affected by telemetry reduction (data were available only for
about $10\%$ of time), so that the lightcurve has very few points (Fig.~\ref{lcurve1}).
Due to the restricted telemetry mode, SPI transmitted to ground only the science housekeeping
data containing the detector's total count rates. In this case, the energy band is $20$~keV -- $8$~MeV,
with a time resolution of $1$~s (Fig.~\ref{lcurve1}).
The IBIS/PICsIT layer, although with a special amount of dedicated telemetry and a reduced number
of channels, was influenced by telemetry gaps (about $62\%$ of events were downloaded),
clearly visible in Fig.~\ref{lcurve}. Moreover, only PICsIT single (i.e. events which deposit their energy in one pixel
only) events were sent to ground,
thus limiting the information for the high energy part of the spectrum.

Considering the duration of the GRB as the time
when the count rate is more than $4\sigma$ above the background rate, we have that GRB021125 was $24$~s
long in the energy range of IBIS/PICsIT, while it is $23$~s according to SPI--ACS.
For IBIS/ISGRI, because of the telemetry gaps, we have only a lower limit of $21$~s.

The IBIS/PICsIT lightcurve (Fig. \ref{lcurve}) in different energy bands shows the indication of a possible softening in the second part of the GRB. This is also consistent with the steep spectrum in the PICsIT energy range (see Sect.~4).
On the other hand, the absence of detection above $500$~keV could be due to the limited PICsIT energy range during this observation.

\begin{table*}
\caption{Sky coordinates of the GRB021125 as seen by the different instruments onboard INTEGRAL and
by the IPN ($3\sigma$).}
\centering
\begin{tabular}{lllc}
\hline
Instrument     & R.A. (J2000)&  Dec. (J2000) & Error radius\\
\hline
IBIS/ISGRI     & 19:47:56    &  +28:23:28    & 2$'$\\
IBIS/PICsIT    & 19:47:51    &  +28:19:16    & 5$'$\\
SPI            & 19:47:55    &  +28:23:49    & 13$'$\\
IPN (centre)   & 19:47:25.93 &  +28:16:0.45  &     \\
IPN (corner 1) & 19:46:49.27 &  +28:09:14.20 &     \\
IPN (corner 2) & 19:47:47.73 &  +28:13:09.68 &     \\
IPN (corner 3) & 19:47:37.78 &  +28:21:06.43 &     \\
IPN (corner 4) & 19:48:36.40 &  +28:25:00.85 &     \\
\hline
\end{tabular}
\label{posizione}
\end{table*}

\section{Imaging}

The fact that PICsIT was in photon--by--photon mode, allowed the extraction of the events in the time region around the burst and the subsequent deconvolution using the standard software IDAS\footnote{\emph{INTEGRAL Data Analysis System}, available at
http://isdc.unige.ch/index.cgi?Soft+download. See also Goldwurm et al. (2003).}.
The same occurred for IBIS/ISGRI, for which the photon--by--photon mode is already the standard operation mode and the only available. In addition, for ISGRI it was possible also to use the IBAS (INTEGRAL Burst Alert System, Mereghetti et al. 2003) off--line software, even though the special set up for PICsIT strongly reduced the available telemetry for ISGRI. Nevertheless, it was possible to obtain images also for ISGRI and to reconstruct the sky position.
Only the scientific housekeeping and on board spectra data were available for SPI, because of the special configuration of INTEGRAL. However, these data allowed to obtain a deconvolved image, although with non--standard techniques.

\begin{figure}
\centering
\includegraphics[scale=0.55]{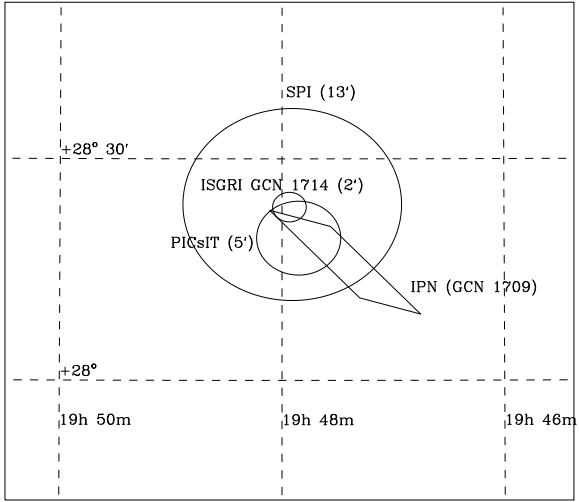}
\caption{Localisation uncertainties  of the GRB021125 from IBIS/ISGRI, IBIS/PICsIT, SPI, and the
Interplanetary Gamma--Ray Burst Timing Network. For more details see Table~\ref{posizione}.}
\label{errorbox}
\end{figure}

The GRB coordinates as seen by INTEGRAL instruments and the IPN are shown in Table~\ref{posizione}. Fig.~\ref{errorbox} shows the error boxes for IBIS/ISGRI (Gros \& Produit 2002), IPN (Hurley et al. 2002), SPI, and IBIS/PICsIT.

The non-standard telemetry configuration of the instruments forced the IBAS\footnote{Normally, IBAS is triggered by the ISGRI layer, but in this case the limited data flow received from the detector telemetry did not allow the triggering.} system to remain in idle mode.
This, together with the fact that INTEGRAL was still in the PV phase, caused a delay in the release of the GRB coordinates of about one day, while the refined error box (with 2 arcmin uncertainty) was released only 3.7 days after the GRB.
Therefore, the fact that no optical counterpart of GRB021125 was found, could mean that the afterglow was below the actual sensitivities of ground telescopes.

\section{Spectral analysis}

IBIS is a coded mask detector and the photons of a single point source are spread all over the detector (e.g. Skinner 2002).
The procedure of spectral extraction for ISGRI consists of the modelling of the illuminated mask by a point
source of unitary flux, placed in the same sky coordinates of the GRB. Then, the model is fitted to the
detected shadowgram in each energy channel to obtain the rate and error for each channel (see Goldwurm et al. 2003 for more
details).
The count spectra obtained by using this procedure, independently implemented in both ISGRI off--line scientific analysis and
IBAS off--line software gave results in very good agreement.

\begin{figure}
\centering
\includegraphics[angle=270,scale=0.36,trim=0 0 0 50,clip]{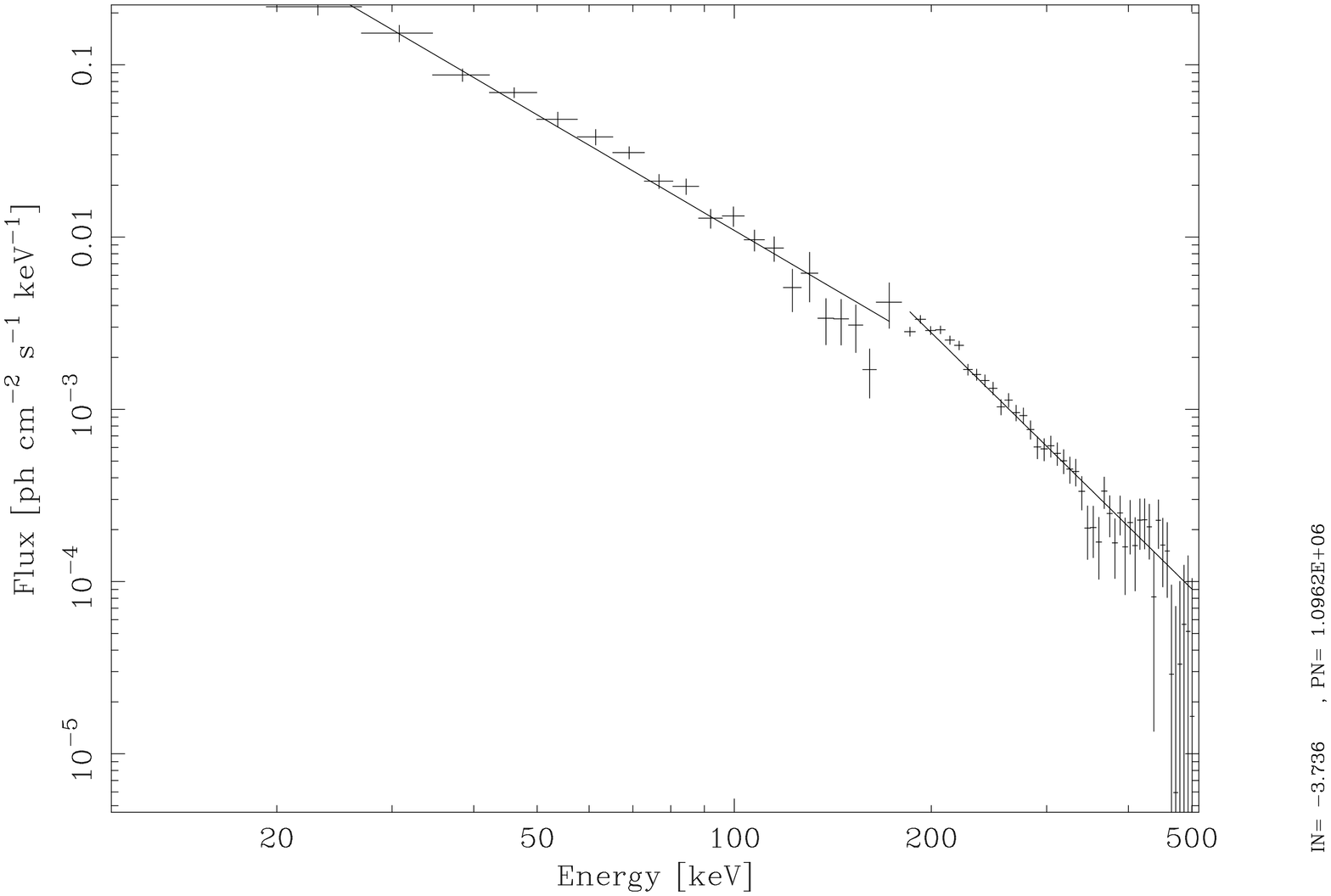}
\caption{IBIS photon spectrum of single events in the energy band $20-500$~keV, fitted with the
two power law models. The boundary between the two layers data is clearly visible.}
\label{spec}
\end{figure}

The high PICsIT count rate during the burst (about $50\%$ more than the background level)
has allowed the use of a more direct procedure. The count spectrum of the GRB has been extracted
by subtracting the background obtained from an empty field observation, and cleaning
for the cosmic--rays induced events. This method implemented both in the Ground Support
Equipement (GSE) and in the Instrument Specific Software (ISSW), has given results in good agreement.

The $20-500$ keV time averaged spectrum from combined data of ISGRI and PICsIT, both corrected for
intrinsic detector deadtimes and telemetry gaps, is shown in Fig.~\ref{spec}. ISGRI refers
to the energy range $20-180$ keV, while PICsIT covers the range $175-500$ keV.
The low energy part of the spectrum (ISGRI) is well fitted with a power law model with $\Gamma\approx 2.2$, while the
PICsIT part is fitted with a power law with $\Gamma\approx 3.7$.
This difference in spectral indices between ISGRI and PICsIT could be the indication
of a spectral break around $200$~keV, but given the present uncertainties in the response of the two layers,
it is not possible to clearly define the energy of the break. In any case, the estimated value would be in agreement
with the statistical distribution of the energy break calculated by Preece et al. (2000) on the basis of about
5500 GRB observed with CGRO/BATSE.
The apparent inconsistency between the ISGRI and PICsIT fluxes at $180-200$~keV,
is to be ascribed to the uncertainty in the ISGRI absolute flux measurement due to
the large data loss caused by the 90\% dead time during the GRB.

The average fluxes, corrected for intrinsic detector deadtimes and telemetry gaps, are
$5.3\pm 0.6$~ph~cm$^{-2}$~s$^{-1}$, and $0.25\pm0.03$~ph~cm$^{-2}$~s$^{-1}$, in the ISGRI and
PICsIT energy bands, respectively. These correspond to $14\pm 2$~Crab for ISGRI and
$9\pm 1$~Crab for PICsIT. The average flux obtained by SPI in the energy range
$0.02-8$~MeV is $9\pm 1$~Crab, in agreement with IBIS. The fluence is approximately
$5.1\times 10^{-5}$~erg~cm$^{-2}$ in the whole IBIS range.

As observed by Ulysses, the GRB had a duration of approximately  30 seconds, a $25-100$~keV
fluence of approximately $8.7\times 10^{-6}$~erg~cm$^{-2}$ and a peak flux of
approximately $7.8\times 10^{-7}$erg~cm$^{-2}$ over $0.50$ seconds (Hurley et al. 2002).

\section{Final remarks}
GRB021125 was the first GRB detected by INTEGRAL in the field of view of the IBIS imager.
The sky coordinates reconstruction with IBIS and SPI are in agreement with each other,
and consistent with the error box of the Interplanetary Network.
The spectrum and lightcurve, obtained with independent methods, which gave consistent results, show an indication
of a possible softening of the spectrum in the second part of the GRB.

GRB021125 has shown the capabilities of the instruments on board the INTEGRAL satellite.

\begin{acknowledgements}
This work has been partially funded by the Italian Space Agency (ASI).
LF acknowledges the hospitality of the INTEGRAL Science Data Centre (ISDC)
during part of this work. AJB acknowledges funding by PPARC grant GR/2002/00446.
\end{acknowledgements}

\end{document}